\documentclass[aps,prl,reprint,groupedaddress]{revtex4-2}

\usepackage{svg}
\usepackage{afterpage}

\usepackage{float}
\usepackage{bigints}
\usepackage{wasysym}
\usepackage{multirow,tabularx}
\newcolumntype{C}{>{\centering\arraybackslash}X}
\usepackage{comment}
\usepackage{cancel}
\usepackage{colortbl}
\usepackage{listings}
\usepackage{amssymb}
\usepackage{physics}
\usepackage{dsfont}
\usepackage{mathtools}
\usepackage{amsmath}
\usepackage{placeins}
\usepackage[compat=1.1.0]{tikz-feynman}
\renewcommand\bra[1]{{\langle{#1}|}}
\renewcommand\ket[1]{{|{#1}\rangle}}

\newcommand{\ee}{\text{e}}
\newcommand{\rr}{\textbf{r}}

\newcommand{\GG}{\textbf{G}}
\newcommand{\qqq}{\textbf{q}}

\newcommand{\kk}{\textbf{k}}
\usepackage{feynmp-auto}
\usepackage{tikz}
\usepackage{diagbox}
\usepackage{standalone}

\usepackage{mathrsfs}
\newcommand{\tadradius}{3mm}
\newcommand{\tadangle}{90}
\tikzset{
tadpole/.style={
particle={},
circle,
minimum size=2*\tadradius,
inner sep=0,
append after command={
\pgfextra
\draw[-{Triangle[width=3pt, length=4pt, sep=0pt -1]}] (\tikzlastnode) +(\tadangle+10:\tadradius) arc[start angle=\tadangle+10, end angle=\tadangle-360, radius=\tadradius];
\endpgfextra
}}}
\begin{document}

%\preprint{}

\title{Accurate Dielectric Response of Solids: Combining the Bethe-Salpeter Equation with the Random Phase Approximation}


\author{Amalie H. Søndersted}
\email[Corresponding author: ]{amhso@dtu.dk}
%\thanks{}
\affiliation{CAMD, Computational Atomic-scale Materials Design, Department of Physics, Technical University of Denmark, 2800 Kgs. Lyngby Denmark}
\author{Mikael Kuisma}
%\thanks{}
\affiliation{CAMD, Computational Atomic-scale Materials Design, Department of Physics, Technical University of Denmark, 2800 Kgs. Lyngby Denmark}
\author{Jakob K. Svaneborg}
%\thanks{}
\affiliation{CAMD, Computational Atomic-scale Materials Design, Department of Physics, Technical University of Denmark, 2800 Kgs. Lyngby Denmark}
\author{Mark Kamper Svendsen}
%\thanks{}
\affiliation{CAMD, Computational Atomic-scale Materials Design, Department of Physics, Technical University of Denmark, 2800 Kgs. Lyngby Denmark}
\author{Kristian S. Thygesen}
%\thanks{}
\affiliation{CAMD, Computational Atomic-scale Materials Design, Department of Physics, Technical University of Denmark, 2800 Kgs. Lyngby Denmark}

\date{\today}

\begin{abstract}
The Bethe-Salpeter equation (BSE) can provide an accurate description of low-energy optical spectra of insulating crystals -- even when excitonic effects are important. However, due to high computational costs it is only possible to include a few bands in the BSE Hamiltonian. As a consequence, the dielectric screening given by the real part of the dielectric function can be significantly underestimated by the BSE. Here we show that universally accurate optical response functions can be obtained by combining a four-point BSE-like equation for the irreducible polarisability with a two-point Dyson equation which includes the higher-lying transitions within the random phase approximation (RPA). The new method is referred to as BSE+. It has a computational cost comparable to the BSE but a much faster convergence with respect to the size of the electron-hole basis. We use the method to calculate refractive indices and electron energy loss spectra for a test set of semiconductors and insulators. In all cases the BSE+ yields excellent agreement with experimental data across a wide frequency range and outperforms both BSE and RPA.  

\end{abstract}

%\keywords{}

\maketitle

% References should be done using th    e \cite, \ref, and \label commands

The current state-of-the-art for calculating optical absorption spectra of solids from first principles is based on many-body perturbation theory\cite{onida2002electronic,onida1995ab,albrecht1998ab,benedict1998theory,rohlfing1998electron,marini2009yambo,yan2012optical}. In a first step, the quasiparticle (QP) band structure is calculated within the GW approximation. In a second step, the two-particle (four-point) Bethe-Salpeter Equation (BSE) is solved within a limited space of electron-hole (e-h) transitions. Because the dimension of the BSE Hamiltonian is $N_c N_v N_k$, where $N_{c/v}$ is the number of conduction/valence bands used to form the e-h basis and $N_k$ is the number of $k$-points, it is only possible to include a few bands close to the Fermi energy. While this approach is sufficient for obtaining a good description of low-energy excitations and the imaginary part of the dielectric function in the corresponding frequency range, it misses the higher-lying excitations and consequently underestimates the real part of the dielectric function even at low frequencies.   

The high computational cost of the BSE approach has motivated attempts to construct exchange-correlation (xc)-kernels ($f_{xc}$) that can account for excitonic effects within the time-dependent DFT\cite{runge1984density} formalism. The advantage of this approach is that the $f_{xc}$ is a two-point function, which renders the Dyson equation much simpler, and allows to include transitions up to very high energies. 

The simplest approximation sets $f_{xc}=0$, which is the random phase approximation (RPA). The RPA can yield a reasonably accurate description of the optical response of bulk metals and semiconductors, but it fails to account for excitons, which are important in systems with weak screening such as insulators and low-dimensional materials\cite{spataru2004excitonic,huser2013dielectric}. 

Attempts to go beyond the RPA with two-point xc-kernels include the parameter free bootstrap kernel\cite{sharma2011bootstrap}, which is based on a postulated (approximate) relation between $f_{xc}$ and the macroscopic dielectric constant $\varepsilon_{\mathrm M}$. The nanoquanta kernel\cite{sottile2003parameter} yields absorption spectra in very good agreement with BSE but at almost the same computational cost. The long-range correction kernel\cite{reining2002excitonic} of the simple form $f_{xc}=-\alpha/q^2$ is computationally efficient but depends on the parameter $\alpha$. 

Thus, the current situation is that one can either solve the four-point BSE to obtain an accurate description of the low-energy excitations but an underestimated screening (due to the neglect of high-energy transitions), or one can use the two-point TDDFT formalism to obtain a better description of screening at the cost of a more approximate description of excitonic effects. 

In this work, we show how the BSE and RPA methods can be combined in a seamless manner to yield a practical scheme for calculating the optical response function of a solid that includes high-energy transitions and at the same time accounts for electron-hole correlations in the low-energy excitations. The new method is referred to as BSE+ and has a computational cost comparable to the BSE. Based on a small test set of semiconductors and insulators, the BSE+ is shown to yield refractive indices and electron energy loss functions in significantly better agreement with experiments than both BSE and RPA.  

To set the stage for the BSE+ method, we first recall the basic equations of the BSE. The excitation energies and associated two-particle wave functions are obtained by solving the eigenvalue equation,
\begin{equation}
    \sum_{S'} \mathcal{H}(\qqq)_{SS'}A^{\lambda}_{S'}(\qqq) = E^{\lambda}(\qqq)A^{\lambda}_{S}(\qqq),
\label{eq:HAEA}
\end{equation}
where $\mathbf q$ is the momentum transfer and $S = \{n,m,\kk\}$ and $S' = \{n',m',\kk'\}$ are indices of the e-h basis. The Hamiltonian matrix takes the form, 
\begin{equation}\label{eq:BSE}
\begin{split}
    \mathcal{H}_{SS'}(\qqq) = (\varepsilon_{m\kk+\qqq} &-  \varepsilon_{n\kk})\delta_{SS'} \\ &- \left(f_{m\kk+\qqq} - f_{n\kk}\right)K_{SS'}(\qqq).
\end{split}
\end{equation}
Here, $f_{n \vb{k}}$ is the occupation function of band $n$ with momentum $\vb{k}$, and the kernel is defined as
\begin{equation}
    K_{SS'}(\qqq) = V^{\mathrm{SR}}_{SS'}(\qqq) - \frac{1}{2}W_{SS'}(\qqq),
\label{eq:kernel}
\end{equation}
where $V^{\mathrm{SR}}_{SS'}$ is the short-range Coulomb interaction (obtained by setting the $\GG=\GG'=0$ component in a plane-wave basis to zero) and $W_{SS'}$ is the screened e-h interaction. The retarded polarisability can be expanded in terms of the solutions to Eq. (\ref{eq:HAEA}), 
\begin{equation}
\begin{split}
     \tilde{P}_{SS'}(\qqq,\omega)= &\sum_{\lambda} A_{S}^{\lambda}(\qqq)A_{S'}^{\lambda}(\qqq)^{*} \\& \times   \left(\frac{f_{m\kk+\qqq} - f_{n\kk}}{\omega-E_{\lambda} +i\eta } - \frac{f_{m'\kk'+\qqq} - f_{n'\kk'}}{\omega+E_{\lambda} + i\eta } \right) ,
\end{split}
\label{eq:BSE_chi_matrix_eq}
\end{equation}
where $\eta$ is a small positive number and we have employed the Tamm-Dancoff approximation under which the eigenvectors $A^{\lambda}(\qqq)$ are orthogonal (note that the $\lambda$-sum runs only over the positive excitation energies). 
Contracting the four-point polarisability to a two-point function and Fourier transforming yields
\begin{equation}
\begin{split}
    \tilde{P}_{\GG\GG'}(\qqq,\omega)= \frac{1}{\Omega} \sum_{SS'}  \rho_{S}(\GG) \tilde{P}_{SS'}(\qqq,\omega) \rho_{S'}(\GG')^*,
\end{split}
\end{equation}
where $\GG$ is a reciprocal lattice vector, $\Omega$ is the crystal volume and
\begin{equation}
     \rho_{S}(\GG)= \rho_{n\kk}^{m\kk+\qqq}(\GG)=\bra{\psi_{n\kk}}\ee^{-i(\qqq+\GG)\rr}\ket{\psi_{m\kk+\qqq}}.
\label{eq:matrix_elements}
\end{equation}
In the optical limit, it can be shown that 
\begin{equation}\label{eq:opticallimit}
\lim_{\mathbf q \to \mathbf 0}\rho_{n\kk}^{m\kk+\qqq}(\mathbf 0)=\frac{i\mathbf q\cdot \langle \psi_{n\kk}|\nabla|\psi_{m\kk}\rangle}{\varepsilon_{n\kk}-\varepsilon_{m\kk}}.
\end{equation}
We have introduced a tilde in Eq. (\ref{eq:BSE_chi_matrix_eq}) to indicate that the sum in practice only runs over a limited set of e-h transitions corresponding to the bands (and $k$-points) used to construct $\mathcal{H}_{SS'}$. We shall denote this set of transitions by $\mathcal T$.

In practice, it is only possible to converge the imaginary part of $\tilde P$ up to a few electron volts above the band gap. It then follows from the Kramers-Kronig relations that the real part cannot be converged even at low frequencies. To address these issues, we introduce a BSE+ polarisability, capturing excitonic effects at the BSE level within the $\mathcal T$ transition manifold while accounting for high-energy transitions at the RPA level, thereby ensuring better convergence of both real and imaginary parts across all frequencies. 

\begin{figure}[H]
    \centering
    \includegraphics[width=0.9\linewidth]{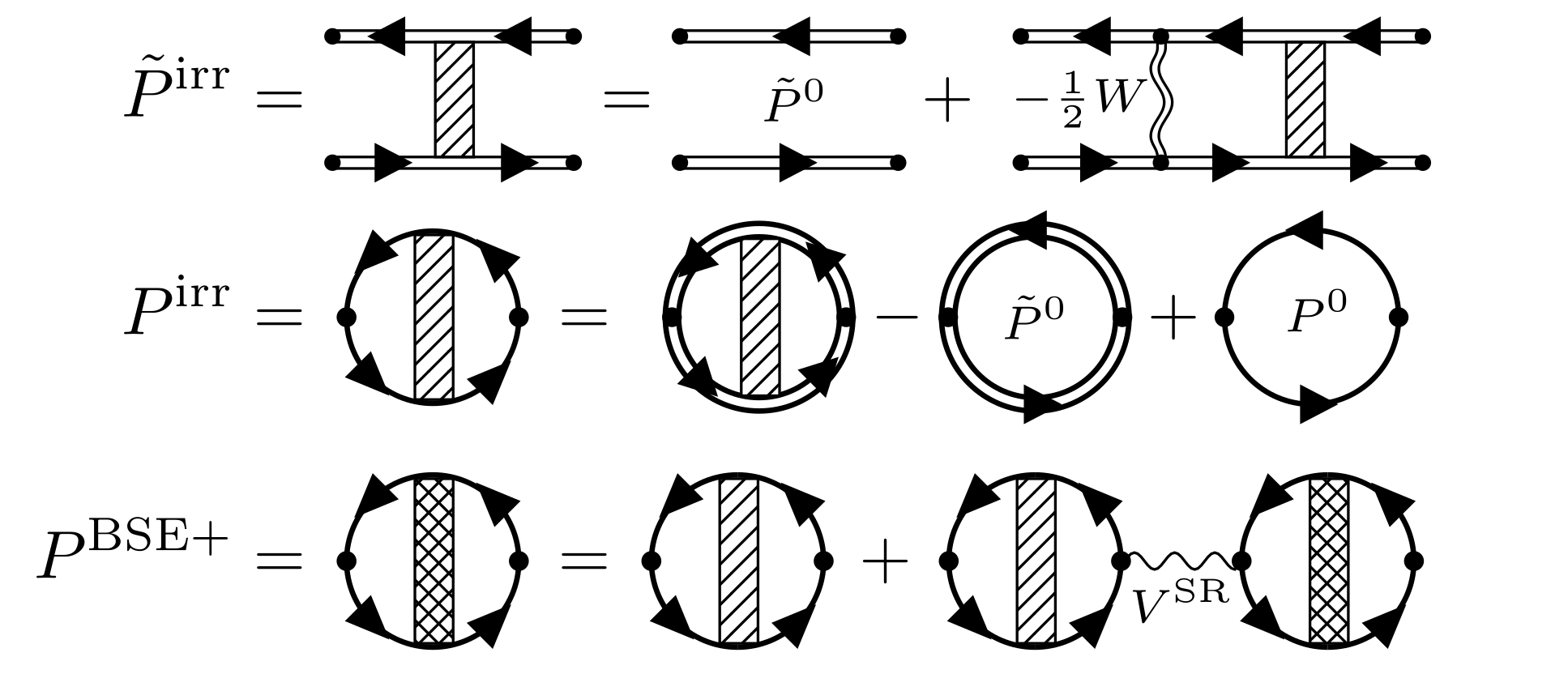}\caption{Feynman diagrams representing Eqs. (\ref{eq:chi_irr}, \ref{eq:chi_irr_new}, \ref{eq:BSE_plus_polarizability}). Double fermion lines represent propagators restricted to the transition space $\mathcal T$, whereas single fermion lines represent full propagators. The double wiggly line represents the screened interaction $W$ and the single wiggly line is the short-range part of the bare Coulomb interaction $V^{\text{SR}}$.}
\label{fig:feynman_diagrams}
\end{figure}

We start by defining an irreducible polarisability $\Tilde{P}^{\text{irr}}$ satisfying the following Dyson equation,
\begin{equation}
\begin{split}
    \Tilde{P}_{SS'}^{\text{irr}}(\qqq,\omega)  &= \Tilde{P}_{SS'}^{0}(\qqq,\omega)  \\ &- \frac{1}{2}\sum_{S_1S_2}\Tilde{P}_{SS_1}^{0}(\qqq,\omega)W_{S_1S_2}(\qqq)\Tilde{P}_{S_2S'}^{\text{irr}}(\qqq,\omega).
\end{split}
\label{eq:chi_irr}
\end{equation}
In practice, this response function is obtained from Eq. (\ref{eq:BSE_chi_matrix_eq}) by solving the eigenvalue equation (\ref{eq:HAEA}) with $V^{\mathrm{SR}}=0$ in Eq. (\ref{eq:kernel}). In the above equation, $\Tilde{P}^{0}$ is the non-interacting (Kohn-Sham) polarisability in the e-h basis and with the sum over bands limited to the transitions $\mathcal T$. The contracted and Fourier transformed $P^{0}$ takes the form
\begin{equation}
\begin{split}
    P_{\GG\GG'}^{0}(\qqq,\omega) = \frac{2}{\Omega}&\sum_{\kk, n,m} (f_{n\kk} - f_{m\kk+\qqq}) \\& \times \frac{\rho_{n\kk}^{m\kk+\qqq}(\GG)\rho_{n\kk}^{m\kk+\qqq}(\GG')^*}{\omega+\varepsilon_{n\kk}- \varepsilon_{m\kk+\qqq}+i\eta}.
\end{split}
\end{equation}
Next, we replace $\tilde P^0$ by $P^0$ (with no constraint on the sum over bands) to obtain  
\begin{equation}
\begin{split}
    P_{\GG\GG'}^{\text{irr}}(\qqq,\omega) &= \Tilde{P}_{\GG\GG'}^{\text{irr}}(\qqq,\omega) - \Tilde{P}_{\GG\GG'}^{0}(\qqq,\omega) + P_{\GG\GG'}^{0}(\qqq,\omega).
\end{split}
\label{eq:chi_irr_new}
\end{equation}
Finally, we obtain the BSE+ result for the polarisability by solving the Dyson equation,
\begin{equation}
\begin{split}
P^{\text{BSE+}}_{\GG\GG'}&(\qqq,\omega) =P_{\GG\GG'}^{\text{irr}}(\qqq,\omega) \\& + \sum_{\GG_1\GG_2} P_{\GG \GG_1}^{\text{irr}}(\qqq,\omega)V^{\mathrm{SR}}_{\GG_1\GG_2}(\qqq)P_{\GG_2\GG'}^{\text{BSE+}}(\qqq,\omega).
\end{split}
\label{eq:BSE_plus_polarizability}
\end{equation}
We note that the BSE polarisability follows from Eq. (\ref{eq:BSE_plus_polarizability}) by replacing $P^{\text{irr}}$ by $\tilde P^{\text{irr}}$. Moreover, by replacing $V^{\mathrm{SR}}$ by the full Coulomb interaction $V$, we obtain the full density-density response function within the BSE+ approximation. Eqs. (\ref{eq:chi_irr}, \ref{eq:chi_irr_new}, \ref{eq:BSE_plus_polarizability}), which constitute the BSE+ method, can be illustrated by the Feynman diagrams in Fig. (\ref{fig:feynman_diagrams}). 

Previous work has stressed the importance of screening the e-h exchange in the BSE by transitions outside the $\mathcal T$-subspace\cite{deilmann2019important,benedict2002screening}. Our work goes beyond the previous work by including such transitions explicitly in the response function.
To evaluate the performance of the BSE+ method, \begin{figure*}[htbp]
\centering
\includegraphics[width=\linewidth]{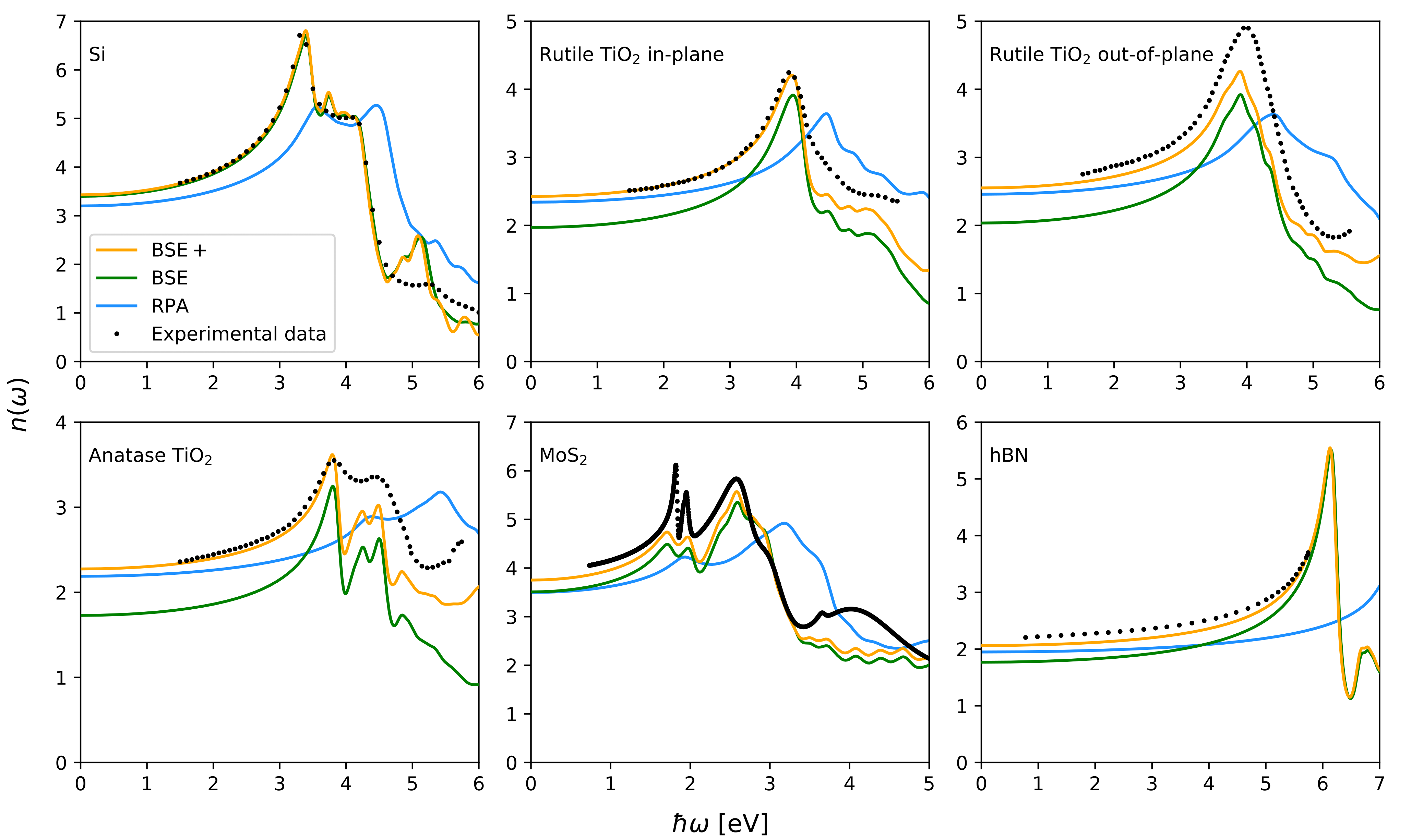}
\caption{Refractive indices calculated with BSE+ (orange), BSE (green), and RPA (blue). The experimental refractive indices are marked with black dots, and are obtained from \cite{Si_experimental_data} (Si), \cite{TiO2_rutile_anatase_exp_data} ($\text{TiO}_2$), and \cite{hBN_refractiveindex} (hBN). For bulk $\text{MoS}_2$, we have obtained the experimental dielectric function from \cite{MoS2_bulk_experimental_eps}, and from this we have calculated the refractive index.}
\label{fig:refractive_indicies}
\end{figure*} we use it to calculate the refractive index of a set of solids and compare it to experimental data and results of the BSE and RPA methods. We have calculated the refractive indices of silicon (Si), molybdenum disulfide ($\text{MoS}_2$), hexagonal boron nitride  (hBN), and titanium dioxide ($\text{TiO}_2$) in both the rutile and anatase phases. The structures were set up using experimental lattice parameters from \cite{Si_lattice_params} (Si), \cite{rutile_lattice_params} (rutile $\text{TiO}_2$), \cite{TiO2_Anatase_latticeParams} (anatase $\text{TiO}_2$), \cite{MoS2_latticeParams} ($\text{MoS}_2$), and \cite{hBN_latticeParams} (hBN). For rutile $\text{TiO}_2$ the refractive index has been calculated for two orientations, namely in-plane and out-of-plane corresponding to the electric field being orthogonal and parallel to the $c$-axis, respectively. For MoS$_2$, hBN, and anatase TiO$_2$ we have only considered the in-plane refractive index. The refractive index depends on both the real and imaginary part of the dielectric function, and it therefore offers a way of evaluating the accuracy of both of these simultaneously. The refractive index is determined from the relation,
\begin{equation}
   n(\omega) = \Re\left( \sqrt{\varepsilon_{\mathrm{M}}(\omega)}\right),
\label{eq:n&eps}
\end{equation}
where $\varepsilon_{\mathrm{M}}(\omega)$ is the macroscopic dielectric function, 
\begin{equation}    
\varepsilon_{\mathrm{M}}(\omega) = 1 - \lim_{\qqq \to \mathbf 0} \frac{4\pi}{q^2}P_{\GG=\GG'=0}(\qqq,\omega).
\label{eq:reciprocal_space_eps}
\end{equation}
The divergence of the Coulomb potential is cancelled in the optical limit by exploiting Eq. (\ref{eq:opticallimit}).

To further evaluate the ability of the BSE+ to describe high frequency responses and plasmonic excitations, we calculate the electron energy loss spectrum (EELS) defined as 
\begin{equation}
    \text{EELS}(\omega) = -\text{Im}\left(1/\varepsilon_{\mathrm M}(\omega)\right).
\label{eq:eels_def}
\end{equation}
We have implemented functions to determine $\tilde{P}^{\text{irr}}$, $\tilde{P}^{0}$ and $P^{0}$ in the BSE and RPA codes in GPAW\cite{enkovaara2010electronic}. For each material, a DFT calculation was performed to determine the ground state wave functions and eigenvalues. Here we used a plane-wave cutoff of 800 eV, the PBE exchange-correlation functional\cite{pbe}, and a $\Gamma$-centered Monkhorst-Pack $k$-point grid with a density of 12 $k$-points per Å$^{-1}$. Subsequently, the ground state density was kept fixed, and energies and wave functions to be used to construct the non-interacting polarisability were calculated on a $k$-point grid with a density of at least 18 per Å$^{-1}$. For the BSE calculations we had to use fewer $k$-points due to memory constraints, resulting in densities between 5 and 8 $k$-points per Å$^{-1}$. A broadening of $\eta = 100$ meV was used for the polarisabilities. The plane-wave cutoff used for $\rho(\mathbf G)$ was in the range of $70-100$ eV for all materials, which was found to be sufficient to account for local field effects. All the computational details can be found in the supplementary information.

For the BSE and BSE+ calculations, we explicitly account for the screened e-h interaction within the transition space $\mathcal T$. We include a given band $n$ in $\mathcal T$, iff $\varepsilon_{n}(\mathbf k)$ lies within $\Delta E_B$ of the valence band maximum or the conduction band minimum for at least one $k$-point.
This ensures that \emph{at least} all transitions up to the energy $E^{\text{QP}}_{\text{gap}}+\Delta E_B$ are included in $\mathcal T$. For all materials we have used $\Delta E_B=2$ eV, except for TiO$_2$ anatase where we have used $\Delta E_B=1.6$ eV due to the larger number of atoms in the unit cell (12 atoms). The corresponding number of bands is listed for each material in the supplementary information.  

As a final comment, PBE is known to underestimate band gaps. As our primary focus has not been on obtaining accurate QP gaps, we have handled this issue by utilizing a scissors operator to adjust the band gaps, aligning the lowest BSE exciton peak with the experimental exciton peak. The fitted band gaps can be found in Table \ref{table:static_n}. The same value for the QP band gap, or equivalently the scissors shift, is used for RPA, BSE, and BSE+. 

The refractive indices of all 5 crystals calculated with BSE+, BSE, and RPA together with experimental data are shown in Fig. \ref{fig:refractive_indicies}. The values of the refractive indices at the lowest energies available in the experimental data, $E_{\text{min}}^{\text{exp.}}$, can be found in Table \ref{table:static_n}.
\begin{table}[h]
\begin{tabular}{|l|l|l|llll|}
\hline
Material                                                                        & $E^{\mathrm{QP}}_{\mathrm{gap}}$ {[}eV{]}&  \multicolumn{1}{l|}{$E_{\text{min}}^{\text{exp.}}$ {[}eV{]}}   & \multicolumn{4}{c|}{$n(\omega)$}                                                                                                  \\ \hline
                                                                                &                             &                      & \multicolumn{1}{l|}{Exp.} & \multicolumn{1}{l|}{BSE+} & \multicolumn{1}{l|}{BSE}  & RPA  \\ \hline
Si                                                                              & 3.35                        & \multicolumn{1}{l|}{1.50}              & \multicolumn{1}{l|}{3.67} & \multicolumn{1}{l|}{3.66} & \multicolumn{1}{l|}{3.63} & 3.36 \\ \hline
\begin{tabular}[c]{@{}l@{}}Rutile $\mathrm{TiO}_2$ \\ in-plane\end{tabular}     & 3.30                        & \multicolumn{1}{l|}{1.49}              & \multicolumn{1}{l|}{2.51} & \multicolumn{1}{l|}{2.51} & \multicolumn{1}{l|}{2.06} & 2.40 \\ \hline
\begin{tabular}[c]{@{}l@{}}Rutile $\mathrm{TiO}_2$ \\ out-of-plane\end{tabular} & 3.30                        & \multicolumn{1}{l|}{1.53}              & \multicolumn{1}{l|}{2.75} & \multicolumn{1}{l|}{2.64} & \multicolumn{1}{l|}{2.13} & 2.52 \\ \hline
Anatase $\mathrm{TiO}_2$                                                        & 4.05                        & \multicolumn{1}{l|}{1.50}              & \multicolumn{1}{l|}{2.36} & \multicolumn{1}{l|}{2.34} & \multicolumn{1}{l|}{1.80} & 2.23 \\ \hline
hBN                                                                             & 7.06                        & \multicolumn{1}{l|}{0.78}              & \multicolumn{1}{l|}{2.20} & \multicolumn{1}{l|}{2.07} & \multicolumn{1}{l|}{1.78} & 1.95 \\ \hline
$\mathrm{MoS}_2$                                                                & 1.87                        & \multicolumn{1}{l|}{0.73}              & \multicolumn{1}{l|}{4.05} & \multicolumn{1}{l|}{3.86} & \multicolumn{1}{l|}{3.61} & 3.56 \\ \hline
\begin{tabular}[c]{@{}l@{}} MAPE \\ from exp. [\%]\end{tabular} &  \multicolumn{3}{l|}{}       & \multicolumn{1}{l|}{2.62} & \multicolumn{1}{l|}{15.87} & 8.36 \\ \hline
\end{tabular}
\caption{Fitted QP band gaps and low-energy refractive indices from experiments and calculations using BSE+, BSE, and RPA, respectively. The experimental data are not available down to the static limit, so we have specified the energy at which the refractive index of each material has been read off. In the last row we show the mean absolute percentage error (MAPE) of the calculated low-energy refractive indices from the experimental values.}
\label{table:static_n}
\end{table} 

With the exception of Si, which is well described by both BSE and BSE+, the BSE underestimates the refractive index across all frequencies. This is due to the neglect of transitions beyond the limited set $\mathcal T$. The RPA misses the excitonic peaks and underestimates the refractive index for frequencies below the band gap. This is due to the neglect of attractive e-h interactions. In all cases, the BSE+ captures the exciton peaks and provides an accurate description of the refractive index over the entire frequency range surpassing both the BSE and RPA in performance with no additional computational cost compared to BSE. We note that the BSE+ presents only little or no improvement over the BSE in terms of predicting the shape and location of the low-energy excitonic peaks. Instead, it provides a better description of the peak height. 
\setcounter{figure}{3}
\begin{figure*}[htbp]
\centering
\includegraphics[width=\linewidth]{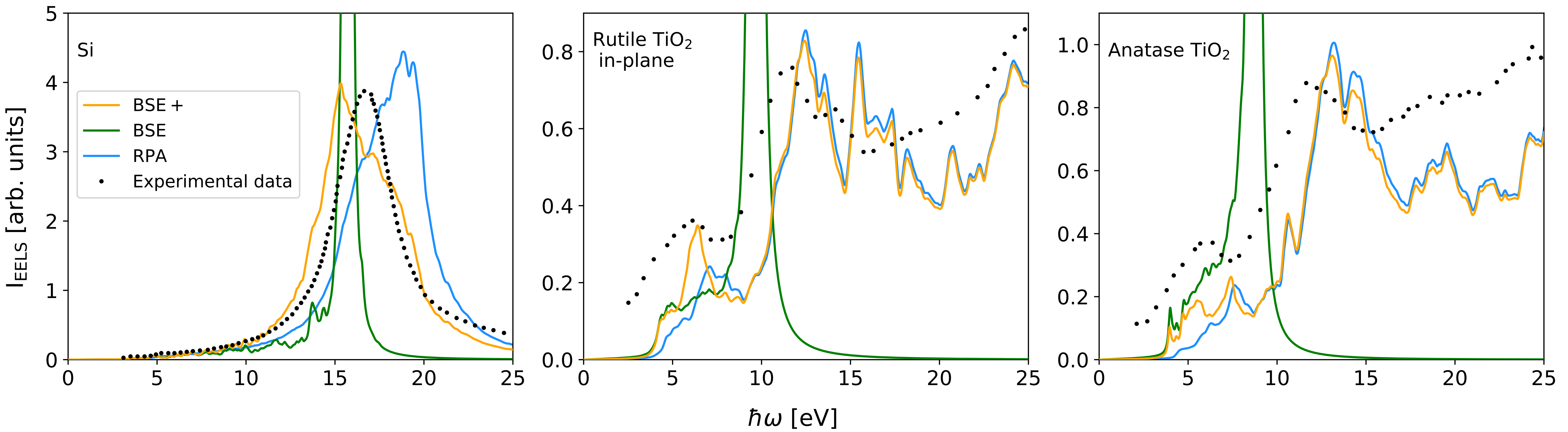}
\caption{EELS spectra calculated with BSE+ (orange), BSE (green), and RPA (blue). The experimental data are marked with black dots and are obtained from \cite{Si_eels} (Si), and \cite{TiO2_rutilex_anatase_eels} ($\text{TiO}_2$). The experimental EELS spectra of the two $\text{TiO}_2$ were multiplied by 35 to approximately match the height of RPA calculated EELS spectra above $\sim$10 eV, which is valid since the unit of the y-axis is arbitrary. }
\label{fig:eels}
\end{figure*}
A key parameter in the theory is the number of bands included in the transition space $\mathcal T$. As the number of bands included in $\mathcal T$ increases, the BSE and BSE+ results will eventually become identical, although the convergence is very slow and not feasible to achieve in practice for all but the simplest materials. Fig. \ref{fig:conv} shows the in-plane refractive index of rutile $\text{TiO}_2$ obtained with BSE and BSE+ as a function of the size of $\mathcal T$. The latter is represented by the parameter $E_T$, denoting the highest energy below which all e-h transitions are included in $\mathcal T$. 
\setcounter{figure}{2}
\begin{figure}[H]
\centering
\includegraphics[width=\linewidth]{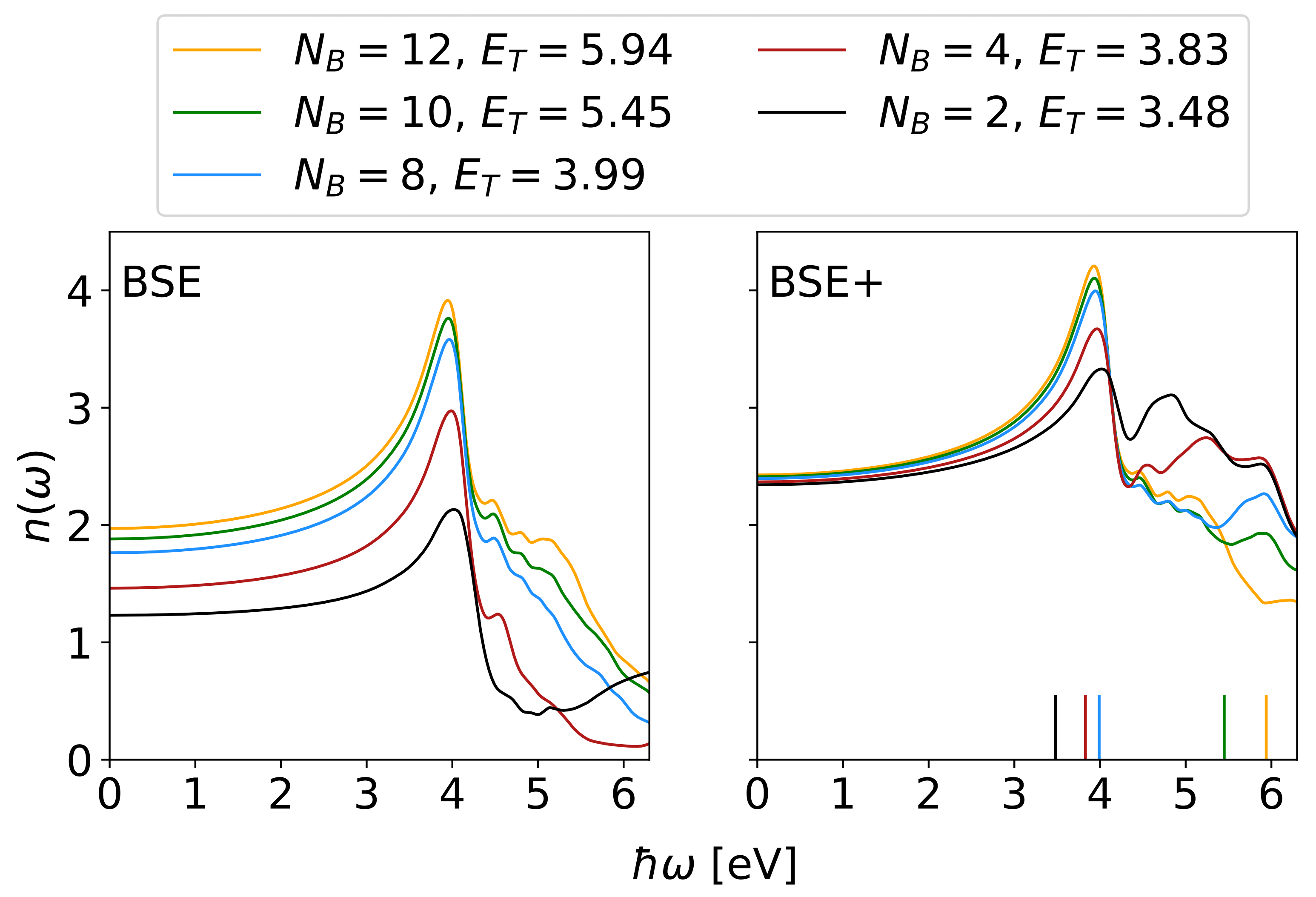}
\caption{Convergence of the in-plane refractive index of rutile $\text{TiO}_2$ with respect to the number of bands included in the BSE and BSE+ calculations.  $N_B$ is the total number of bands in the BSE/BSE+ calculations divided equally between valence and conduction bands. The parameter $E_T$ denotes the energy below which all e-h transitions are included given the number of bands in the calculation, and the values are marked in the BSE+ plot with vertical lines.}
\label{fig:conv}
\end{figure} 
From Fig. \ref{fig:conv} we see that the low-energy refractive index from BSE converges very slowly with the number of bands, while BSE+ yields a converged value already with two bands. The faster convergence of the BSE+ calculations with the number of bands stems from the following: While BSE completely neglects all transitions beyond $\mathcal T$, the BSE+ merely neglects the e-h attraction ($W$) in the transitions beyond $\mathcal T$, which is obviously a much more gentle approximation. It can be seen that the refractive index from BSE+ generally is well converged up to the transition energy threshold $E_T$, which provides a simple means to estimate and control the energy range in which a BSE+ calculation can be expected to be converged. 

We now turn to the description of electron energy loss spectra (EELS). In Fig. \ref{fig:eels} we show the $q=0$ EELS of Si and TiO$_2$ in the rutile and anatase phases as calculated with RPA, BSE, BSE+, and compared to experimental data. In all cases, the BSE falls short due to the lack of high-energy excitations. As also seen for the refractive indices, the RPA-based EELS do not capture the excitonic features in the low-energy range. On the other hand, the BSE+ reproduces both the gross features in the experimental spectra, which are mainly governed by plasmonic excitations, and the finer structures of excitonic origin around the band edge. In the case of Si, the e-h attraction does not produce distinct excitonic peaks, but red-shifts the large plasmon peak (as compared to the RPA result). This shift is captured by the BSE+ despite the plasmon energy ($\sim$17 eV) being much larger than the threshold energy of 2 eV used to select the bands to be included in $\tilde P^{\mathrm{irr}}$. For anatase and rutile $\text{TiO}_2$ the BSE+ result is close to the RPA (and experiments) for energies above 10 eV, while the excitonic peaks around 4-7 eV are better described by BSE+.

In conclusion, we have introduced the BSE+ method, which extends the well known BSE method by including transitions outside the active BSE e-h space. The additional transitions are included at the RPA level, which makes it possible to complete the transition space without increasing the computational cost. We have shown that relative to standard BSE and RPA, the BSE+ method significantly improves the description of dielectric screening, refractive indices, and electron energy loss spectra because it accounts for excitons and plasmons simultaneously. The method can be further extended by using any two-point TDDFT kernel to account for xc-effects among the additional transitions. The method could also form the basis for total energy calculations based on the adiabatic connection fluctuation dissipation theorem (ACFDT), where a response function with an accurate description of low-energy excitations and inclusion of high-energy transitions is of key importance\cite{olsen2014static}.

\begin{acknowledgments}
We acknowledge funding from the European Research Council (ERC) under the European Union’s Horizon 2020 research and innovation program Grant No. 773122 (LIMA) and Grant agreement No. 951786 (NOMAD CoE). K. S. T. is a Villum Investigator supported by VILLUM FONDEN (grant no. 37789).
\end{acknowledgments}

\bibliography{bibliography}

\end{document}